\documentclass[12pt]{article}
\usepackage{amsfonts}
\usepackage{amsmath}

\newcommand{\newsection}{    
\setcounter{equation}{0}\section}
\def\appendix#1{\addtocounter{section}{1}\setcounter{equation}{0}
\renewcommand{\thesection}{\Alph{section}}
\section*{Appendix \thesection\protect\indent \parbox[t]{11.15cm}{#1}}
\addcontentsline{toc}{section}{Appendix \thesection\ \ \ #1}}

\newcommand{\be}{\begin{eqnarray}}
\newcommand{\ee}{\end{eqnarray}}
\newcommand{\bea}{\begin{eqnarray}}
\newcommand{\eea}{\end{eqnarray}}
\newcommand{\ba}{\begin{array}}
\newcommand{\ea}{\end{array}}
\newcommand{\nn}{\nonumber \\}

\def \la {\label}

\newcommand{\lc}{\lrcorner}
\def\a{\alpha}
\def\b{\beta}

\def\cL{{\cal L}}

\def\bo{{\bar{1}}}

\def\bbe{{\bf{e}}}
\font\mybb=msbm10 at 11pt

\def\bb#1{\hbox{\mybb#1}}

\def\bR {\bb{R}}

\def\cD {{\cal{D}}}

\def\hn {{\tilde{\nabla}}}

\begin{document}
\begin{titlepage}
\begin{center}
\vspace{5.0cm}

\vspace{3.0cm} {\Large \bf Topology of supersymmetric ${\cal N}=1, D=4$ supergravity horizons}
\\
[.2cm]

\vspace{1.5cm}
 {\large  J. Gutowski and  G. Papadopoulos}

\vspace{1cm}
 Department of Mathematics\\
King's College London\\
Strand\\
London WC2R 2LS, UK\\

\vspace{0.5cm}

\end{center}

\vskip 1.5 cm

\begin{abstract}
All supersymmetric ${\cal N}=1, D=4$ supergravity horizons have toroidal or spherical topology,
irrespective of whether the black hole preserves any supersymmetry.

\end{abstract}
\end{titlepage}


\newsection{Introduction}

Supersymmetric  phenomenological models are based on
 ${\cal N}=1, D=4$  supergravity.  This theory has 4 supersymmetries, and the constraints imposed
 on the couplings are rather weak, see e.g. \cite{wess} and references within. Therefore a large class of models can exist even for a prescribed
 matter content.  Nevertheless, ${\cal N}=1, D=4$  supergravity  exhibits properties which are independent of
 the particular chosen model. Such properties characterize  the theory.
 One such property which we shall investigate here is the topology of supersymmetric horizons of extreme,
  but not necessarily supersymmetric,
 black holes of ${\cal N}=1, D=4$  supergravity.  In particular we show that for any matter content, and under some mild
 restrictions which we explain later,  the associated horizon sections
   have topology\footnote{To our knowledge the existence of black holes of ${\cal N}=1, D=4$
   supergravity with $T^2$ horizon
   topology has not been ruled out, however see \cite{hawking}, \cite{chrusc}.}
     $T^2$ or $S^2$. In addition, the metric in the spherical case is a product $\bR^{1,1}\times S^2$.
     To prove this, we use the classification of supersymmetric backgrounds of
     ${\cal N}=1, D=4$  supergravity
   \cite{class2, class3}, and the fact that horizon sections are  connected and closed  Riemann
   surfaces which are characterized  by their Euler number. An explicit example
     of a spherical horizon has been given in \cite{mo}.

Black hole horizons in 4-dimensions have been extensively investigated following the
uniqueness theorems in \cite{israel} -\cite{robinson}. More
recently, attention has been focused on the near horizon geometries of black holes
associated with an (ungauged) Einstein-Maxwell system coupled
to scalars that includes a potential term. In particular,  it can be shown under some mild
assumptions that the near horizon geometry of such
a black hole has an $O(2,1)$ symmetry. This follows from the results of \cite{kunduri} and
\cite{hollands}. The action of ${\cal N}=1, D=4$
supergravity differs from the above Einstein-Maxwell system in two respects. First the gauge
fields are allowed to be non-abelian and second the
scalars are gauged. However, we make the additional assumption that the near
horizon geometry is supersymmetric.

Before we proceed with the analysis, we shall first put our result into context.
For appropriately chosen couplings and
matter content most of the well-known 4-dimensional black holes, like the Schwarzschild, Reissner-Nordstr\"om, Kerr,  and Kerr-Newman,
 can be embedded as solutions
 of ${\cal N}=1, D=4$  supergravity. From these black holes only those with extreme horizons
 are of relevance here. Even in this case,  near horizon geometries, like that of the Reissner-Nordstr\"om black hole,
 are not supersymmetric in  ${\cal{N}}=1$ supersymmetry,
 suggesting that supersymmetric horizons may not exist. Nevertheless this argument is not conclusive for restricting the existence
 of supersymmetric horizons, as it applies only to a particular class of solutions.
More generally, one may use  energy bounds \cite{schoen, witten, gibbons} and observe that in the construction of a Nester tensor the  standard gravitino
 supercovariant connection\footnote{In the energy bound of \cite{gibbons} the supercovariant connection
 of simple ${\cal N}=2, D=4$ supergravity was used which includes a Maxwell field.} of ${\cal N}=1, D=4$  supergravity does not include a Maxwell field.
 As a result no electric or magnetic
  charge can be detected
 at asymptotic infinity indicating that there are no supersymmetric charged black holes. However to establish
 such bounds one needs
 at least the weak energy condition which does not hold for all matter couplings of ${\cal N}=1, D=4$
 supergravity. Moreover, we do not assume
 that the black hole spacetime is supersymmetric or put any conditions on the asymptotic geometry
 allowing for example for $AdS_4$ black holes.
 Thus although our assumption that the near horizon geometry is
 supersymmetric is rather restrictive, the indirect arguments provided above for the existence of
 such geometries are not conclusive.
 The advantage of our approach is that it has a wide range of applicability, which includes any matter content,
 subject to some assumptions
 which we now explain. There are three types of restrictions required for the technical proof.
 First,   the kinetic terms of the gauge fields and scalars are canonical, ie the gauge group metric and the K\"ahler metric
 of the scalars  are positive definite. Second, if the conditions
 imply that the gauge field vanishes, then the gauge potential is always chosen to be the trivial one, ie
 we do not consider the cases for which there are flat but  non-trivial connections. Third, at several places
 we have assumed that  fields and tensors are sufficiently smooth and sometimes analytic. This restriction is mentioned
 as it arises in the proof.

 In the considerations that follow, we do not assume that the black hole spacetime is supersymmetric.
 We only assume that the near horizon geometry is. This is an important distinction as there exist non-supersymmetric
 black holes with supersymmetric horizons \cite{nonbps1, nonbps2}.  To implement this in our analysis, we distinguish between the stationary Killing vectors
 which belong to the equivalence class that characterize the horizon\footnote{This class includes the stationary Killing vector field of a black hole,
 see eg \cite{hethor} for details.}  of a
  black hole
  and the Killing vector field
 constructed as Killing spinor bilinear. A similar approach has recently been taken in the context of
``pseudo-supersymmetric" extremal near-horizon solutions
of the minimal de-Sitter five-dimensional supergravity theory \cite{desit}.
In this theory, the vector field obtained as a Killing spinor bilinear is not Killing, and so one cannot identify this bilinear with the
stationary Killing vector of the black hole.
This differs from previous analysis done in the context of 5-dimensional
 black holes in \cite{reallbh}, where the two were identified, and so it is assumed that both the
 near horizon geometry and the black hole spacetime are supersymmetric.

The plan of the paper is as follows. In section 2, we describe in greater detail the assumptions we make,
and the construction of a basis adapted to the Gaussian Null co-ordinates in the near-horizon limit.
We also analyse solutions of the Killing spinor equations (KSEs) corresponding to supersymmetric
extremal near-horizon geometries in minimal ${\cal{N}}=1$, $D=4$ supergravity.
In section 3 we use this analysis to prove that the event horizon must have a toroidal topology.

\newsection{Horizons}

\subsection{Supersymmetric horizons of non-supersymmetric black holes}

A starting point in the analysis of near-horizon geometries in the context of supersymmetry is the
identification of the stationary Killing vector field of a black hole with a Killing vector field constructed as a Killing spinor bi-linear \cite{reallbh}.
In such an investigation it is assumed that the Killing spinor bilinear near the horizon can be extended to a Killing vector on the spacetime.
Both near horizon geometry and black hole spacetime are supersymmetric.

Here, we  address a different situation, where one has an extremal black hole which does not necessarily preserve
any supersymmetry outside the horizon, but whose near-horizon geometry is supersymmetric. In particular, we do not assume that
 the Killing spinor bilinears
of the near horizon geometry can  be extended into the bulk spacetime. Moreover, it no longer follows  that the
horizon is a Killing horizon of a Killing spinor bilinear vector field.
We remark that for a number of theories, with various asymptotic conditions, it has been
shown that black hole event horizons are Killing horizons,
\cite{hawking}, \cite{hawkingellis}, \cite{chrusc2}, \cite{hollands1}, \cite{monc}, \cite{hollands}.
However, this result has not yet been established for the generic ${\cal{N}}=1$, $D=4$ supergravity theory we consider here,
so we shall  simply assume that the event horizons of the black holes we consider are Killing horizons.

Adapting Gaussian null coordinates \cite{gnull} with respect to a stationary Killing vector field of an extreme black hole
and taking the near horizon limit, the bosonic fields of the ${\cal N}=1, D=4$ supergravity can be written as
\be
\label{gn1}
ds^2 &=& 2 du (dr+r h_I dy^I -{1 \over 2} r^2 \Delta du) + \tilde g_{IJ} dy^I dy^J~,
\cr
A^a&=& r \Phi^a du + B^a_I dy^I~,~~~~\phi^\alpha=\phi^\alpha(y)~,
\ee
where the components of the 4-dimensional spacetime metric $ds^2$, $\Delta$,  $h$ and  $\tilde g$ are independent of $u, r$, and similarly
for the components $\Phi$ and $B$ of the gauge potential $A$. The scalar fields $\phi$ depend only on $y$. The horizon section
${\cal S}$ is given by the co-dimension two subspace $r=0, u={\rm const}$ and it is required to be oriented, connected, compact without boundary.
The metric on ${\cal S}$ is $\tilde g$. Observe that the horizon is Killing with respect to $\partial_u$.
However, in the analysis that will follow $\partial_u$ will not be identified with a Killing spinor bi-linear.

It  is most convenient to introduce  a particular basis adapted to the Gaussian null co-ordinates. This also enables
one to simplify the solution of the KSEs and to make optimal use of compactness of spatial cross sections of the horizon. The latter
requirement significantly constrains the spacetime geometry.
The basis we shall use is given by

\be
\label{basis1}
\bbe^+ = du, \qquad \bbe^- = dr + rh -{1 \over 2} r^2 \Delta du, \qquad \bbe^i = e^i_I dy^I~,~~~i,j=1,3~,
\ee
and the metric and gauge potential can be rewritten as
\be
ds^2 &=&2 \bbe^+ \bbe^- + \delta_{ij} \bbe^i \bbe^j
\cr
A^a&=& r \Phi^a \bbe^+ + B^a_i \bbe^i~,~~~~\phi^\alpha=\phi^\alpha(y)~.~
\ee
The components of the spin connection associated with this basis are presented in Appendix A.

\subsection{Killing spinor equations}

Since we assume that the near horizon geometries are supersymmetric, the horizons must be solutions
of the KSEs of ${\cal N}=1, D=4$ supergravity. These equations have been solved in all generality
\cite{class2, class3}. However, these results are not directly applicable in our case as the natural frame associated with
near horizon geometries is different from that adapted to supersymmetric solutions \cite{class2}. So to distinguish which of
the supersymmetric solutions are near horizon geometries
some of the analysis must be repeated. Moreover, we have to impose that ${\cal S}$ is compact and this condition
is not included in the investigation of supersymmetric solutions.

The action and supersymmetry transformations of ${\cal N}=1, D=4$ supergravity that we shall use
can be found in  \cite{wess} and references within. Our notation follows that of \cite{class2} and \cite{wess}, with some minor changes.
In particular, the gravitino KSE is
\be
\label{d4grav}
\nabla_\mu \epsilon_L + V_\mu \epsilon_L +{i \over 2} e^{K \over 2} W \gamma_\mu \epsilon_R =0~,
\ee
the gaugino KSE is
\be
\label{gaugino}
F^a_{\mu \nu} \gamma^{\mu \nu} \epsilon_L -2i \mu^a \epsilon_L =0~,
\ee
and the KSE associated with the chiral multiplets is
\be
\label{scal}
i \cD_\mu \phi^\alpha \gamma^\mu \epsilon_R - e^{K \over 2} G^{\alpha \bar{\beta}} D_{\bar{\beta}} {\bar{W}} \epsilon_L=0~.
\ee
Here $\nabla$ is the  spin connection of 4-dimensional spacetime,
$K=K(\phi^\alpha, \phi^{\bar{\beta}})$ and $G_{\alpha \bar{\beta}}= \partial_\alpha \partial_{\bar{\beta}} K$ are
 the K\"ahler potential and K\"ahler metric of the sigma model manifold $S$, respectively,
and $W=W(\phi^\alpha)$ is
a holomorphic potential. Moreover,
\be
D_\alpha W = \partial_\alpha W + \partial_\alpha K W,
\qquad
\cD_\mu \phi^\alpha = \partial_\mu \phi^\alpha - A^a_\mu \xi^\alpha_a
\ee
where $\xi_a$ are holomorphic Killing vector fields on  $S$.
\be
F^a = d A^a - f^a{}_{bc} A^b \wedge A^c~,
\ee
is the field strength of the gauge potential $A$ which arises from gauging isometries in $S$,
where $f^a{}_{bc}$ are the structure constants. $\mu_a$ is the moment map defined by
\be
G_{\alpha \bar{\beta}} \xi^{\bar{\beta}}_a =  i \partial_\alpha \mu_a~.
\ee
The 1-form $V$ which enters into the gravitino KSE ({\ref{d4grav}}) is
given by
\be
V_ \mu = {1 \over 4} \big( \partial_\alpha K \cD_\mu \phi^\alpha
- \partial_{\bar{\alpha}} K \cD_\mu \phi^{\bar{\alpha}} \big)~.
\ee

The spinors $\epsilon_R, \epsilon_L$ are chiral spinors satisfying
\be
\gamma_5 \epsilon_L = \epsilon_L, \qquad \gamma_5 \epsilon_R = - \epsilon_R
\ee
where $\gamma_5 = i \gamma_{0123}$. $\epsilon_R$, $\epsilon_L$ are related by
\be
\epsilon_R = C * \epsilon_L
\ee
where $C= -\gamma_{012}$ is the charge conjugation matrix.
We will find it convenient to decompose spinors as
\be
\epsilon_L = \epsilon_{L+} + \epsilon_{L-}, \qquad \epsilon_R = \epsilon_{R+} + \epsilon_{R-}~,
\ee
where
\be
\gamma_+ \epsilon_{L+}= \gamma_+ \epsilon_{R+} = \gamma_- \epsilon_{L-}=\gamma_- \epsilon_{R-}=0~,~~~\gamma_{\pm}={\pm \gamma_0+\gamma_2\over \sqrt{2}}~.
\ee

\subsection{Solution of KSEs}

To solve the KSEs we first decompose them along the light-cone and transverse directions and use the analyticity
of the fields in the $r$ and $u$ coordinates. As a general
rule the conditions which arise from the light-cone directions can be solved directly. This together
with elements from spinorial geometry \cite{spingeom} as well as the compactness of ${\cal S}$ allows us to show that
${\cal S}$  is topologically $T^2$.

\subsubsection{Gaugino}

The components of the gauge field strength are
\bea
F^a_{+-} &=& - \Phi^a, \qquad F^a_{+i} = r \big(-\partial_i \Phi^a + \Phi^a h_i -2f^a_{bc} \Phi^b B^c_i\big)~,
\qquad
\cr
F^a_{ij} &=& (dB^a)_{ij}-2 f_{bc}^a\, B^b_i\, B^c_j~.
\eea
On substituting into the gaugino KSE ({\ref{gaugino}}), one obtains
\be
\label{exv1}
\big(  \Phi^a -i F^a_{12} - i \mu^a \big) \epsilon_{L+} =0~,
\ee
and
\be
\label{exv2}
\big( - \Phi^a +i F^a_{12} - i \mu^a \big) \epsilon_{L-} +r \big(-\partial_i \Phi^a + h_i \Phi^a -2f^a_{bc} \Phi^b B^c_i\big) \gamma_-
\gamma^i \epsilon_{L+} =0
\ee
It is clear that in order for these equations to admit solutions other than $\epsilon_L=0$, one must have
\be
\label{magn}
\Phi^a=0~.
\ee
Thus the remaining equations are
\bea
(\mp  F^a_{12} -  \mu^a \big) \epsilon_{L\pm}=0~.
\eea
If both $\epsilon_{L\pm}\not=0$, then
\bea
F^a=0~,~~~\mu^a=0~.
\la{famao}
\eea
Note that this does not mean that $\mu^a$ vanishes identically. It simply vanishes on the space of solutions.

\subsubsection{Gravitino}

Using  (\ref{magn}), we first integrate the  $+$ and $-$ components of ({\ref{d4grav}})  to find
\be
\epsilon_{L+} = \eta_{L+}, \qquad
\epsilon_{L-} = r \big({1 \over 4} h_i \gamma_- \gamma^i \eta_{L+}-{i \over 2}
e^{K \over 2} W \gamma_- C* \eta_{L+} \big) + \eta_{L-}~,
\ee
where $\eta_{L\pm}$ do not depend on $r$. These in turn are   given by
\be
\eta_{L-} = \tau_{L-}, \qquad
\eta_{L+} = u \big({1 \over 4} h_i \gamma_+ \gamma^i \tau_{L-} -{i \over 2} e^{K \over 2} W \gamma_+
C* \tau_{L-} \big) + \tau_{L+}~,
\ee
where $\tau_{L\pm}$ do not depend on $r$ and $u$. One also finds the  conditions
\be
\label{alg1}
\Delta +{1 \over 4} h^2 -e^K |W|^2 =0~,
\ee
\be
\label{alg1b}
dh=0~,
\ee
and
\be
\label{alg2}
 \Delta h_i - \partial_i \Delta=0~.
\ee
The latter is a parallel transport equation for $\Delta$. As a result,  $\Delta$ is either {\it positive} or {\it negative}. Moreover if $\Delta$
vanishes at a point, it vanishes everywhere on ${\cal S}$.

Next consider the remaining components of ({\ref{d4grav}}), these imply that
\be
\label{covdx}
\hn_i \eta_{L\pm} \mp {1 \over 4} h_i \eta_{L \pm} + V_i \eta_{L \pm}
+{i \over 2} e^{K \over 2} W \gamma_i C* \eta_{L \pm} =0~,
\ee
where $\hn_i$ denotes the spin connection on the  horizon section ${\cal S}$.
One also obtains
\be
\label{alg3}
\big( {1 \over 4} \hn_j h_i \gamma^j -{1 \over 8} h_i h_j \gamma^j
+{1 \over 2} e^K |W|^2 \gamma_i \big) \eta_{L+}
- \big({i \over 2} \hn_i (e^{K \over 2} W) + iV_i e^{K \over 2} W \big) C* \eta_{L+}=0~.
\ee

Writing the above conditions in terms of $\tau_{L \pm}$,
 one finds

 \be
\label{covdxx}
\hn_i \tau_{L\pm} \mp {1 \over 4} h_i \tau_{L \pm} + V_i \tau_{L \pm}
+{i \over 2} e^{K \over 2} W \gamma_i C* \tau_{L \pm} =0~,
\ee
and
\be
\label{alg3x}
\big( {1 \over 4} \hn_j h_i \gamma^j -{1 \over 8} h_i h_j \gamma^j
+{1 \over 2} e^K |W|^2 \gamma_i \big) \tau_{L\pm}
- \big({i \over 2} \hn_i (e^{K \over 2} W) + iV_i e^{K \over 2} W \big) C* \tau_{L\pm}=0~.
\ee

The integrability conditions of either (\ref{covdx}) or (\ref{covdxx}) imply that
\bea
\pm \tilde R_{{\cal S}}\tau_{L\mp}=4i \epsilon^{ij} \tilde \nabla_{[i}V_{j]} \tau_{L\mp}\pm 2\Delta \tau_{L\mp}\pm \tilde\nabla^k h_k \tau_{L\mp}~,
\la{incon}
\eea
where $\tilde R_{{\cal S}}$ is the Ricci scalar of the horizon section ${\cal S}$.

Returning to the gaugino KSE ({\ref{gaugino}}) gives, in addition to ({\ref{magn}}),
the following algebraic conditions
\be
\label{alg4}
\big(\mu^a \pm F^a_{12}\big) \tau_{L\pm}=0~,
\ee
and
\be
\label{alg5}
\mu^a \big( {1 \over 4} h_i \gamma^i \eta_{L+} -{i \over 2} e^{K \over 2} W C* \eta_{L+} \big) =0~.
\ee
Observe that if $\tau_{L\pm}\not=0$, then $F^a=\mu^a=0$ as in (\ref{famao}).

\subsubsection{Chiral}

{}For completeness, the chiral KSEs ({\ref{scal}})  imply
\be
\label{alg6}
i \cD_j \phi^\alpha \gamma^j C* \eta_{L\pm} - e^{K \over 2} G^{\alpha \bar{\beta}}
D_{\bar{\beta}} {\bar{W}} \eta_{L\pm} =0~,
\ee
and
\be
\label{alg7}
e^{K \over 2} {\bar{W}} \cD_j \phi^\alpha \gamma^j \eta_{L+}
+\big(-i h^j \cD_j \phi^\alpha +i e^K W G^{\alpha \bar{\beta}} D_{\bar{\beta}} {\bar{W}} \big)
C* \eta_{L+}=0~.
\ee

\subsubsection{Killing vector bi-linear}

The Killing spinor $\epsilon$ of the near horizon geometry is associated with a null 1-form $Z$ which in turn gives rise
to a Killing vector.
From the results of {\cite{class2}}, it is known that
\be
i_Z F^a=0, \qquad i_Z \cD \phi^\alpha =0~,
\la{susycona}
\ee
where we have used $Z$ to denote both the 1-form and the associated vector. To continue, it is useful to compute $Z$.
For this, we set
\be
\tau_{L+}= a\,1, \qquad \tau_{L-}= b\, e_{12}
\la{aba}
\ee
where $a, b$ are $r,u$-independent complex functions.
We also adopt a Hermitian basis $\bbe^1, \bbe^\bo$ for ${\cal{S}}$, with respect to which
\be
\gamma_1 = \sqrt{2}\, e_1 \wedge, \qquad \gamma_\bo = \sqrt{2} \,{e_1}\lc~.
\ee
The Killing spinor can be rewritten as
\bea
\epsilon_{L+}&=& \big(a + u({1 \over 2} b h_1 +{i \over \sqrt{2}} e^{K \over 2} W {\bar{b}}) \big)1
\nn
\epsilon_{L-} &=&\big((1+{1 \over 2} \Delta u r) b + r(-{1 \over 2} a h_\bo +{i \over \sqrt{2}} e^{K \over 2}
W {\bar{a}}) \big) e_{12}~.
\eea
The 1-form bilinear can be computed using
\be
Z = \langle \Gamma_{12} \epsilon^*, \gamma_A \epsilon \rangle\,\bbe^A~,
\ee
where $\langle\cdot,\cdot\rangle$ is the standard Hermitian inner product. An explicit expression is given in
appendix A.
We have assumed that the near horizon geometry is supersymmetric. We furthermore assume that all
gauge-invariant spinor bilinears constructed from the Killing spinor $\epsilon$ are smooth and well-defined
everywhere on the near-horizon geometry.

\subsubsection{Einstein Equation}

The  Einstein equation is
\bea
&&R_{MN} - 2 G_{\alpha \bar{\beta}} \cD_{(M} \phi^{\alpha}
\cD_{N)} \phi^{\bar{\beta}} -{\rm Re} (H_{ab}) F^a{}_{ML} F^b{}_N{}^L
+ 2g_{MN} \big({1 \over 8}  {\rm Re} (H_{ab})  F^a{}_{L_1 L_2} F^{b L_1 L_2}
\cr
&&\qquad
-{1 \over 4} \mu_a \mu^a -{1 \over 2} e^K (G^{\alpha \bar{\beta}}D_\alpha W
D_{\bar{\beta}} {\bar{W}} -3 |W|^2) \big)=0~.
\eea
From the ``+-" component, one has
\be
\label{alge2}
{1 \over 4} \hn^i h_i -{1 \over 4} h^2 -{1 \over 2} \Delta +{3 \over 2} e^K |W|^2
-{1 \over 2} e^K G^{\alpha \bar{\beta}} D_\alpha W D_{\bar{\beta}} \bar{W} =0~.
\ee
If $\Delta \neq 0$, then on using (\ref{alg2}) and ({\ref{alg1}}), one finds that
({\ref{alge2}}) is equivalent to
\be
\label{alge3}
{1 \over 4} \hn^2 \Delta -{1 \over 8} \Delta^{-1} \hn_i \Delta \hn^i \Delta
= - \Delta^2 +{1 \over 2} \Delta e^K  G^{\alpha \bar{\beta}} D_\alpha W D_{\bar{\beta}} \bar{W}~.
\ee
It then follows, as a consequence of the maximum principle, and compactness of
${\cal{S}}$ that one cannot have $\Delta<0$. So one must have
\bea
\Delta\geq 0~.
\eea

The $ij$ components of the Einstein equations imply that
\be
\tilde R_{{\cal S}} - 2 \mu_a \mu^a -2 G_{\alpha \bar{\beta}} \cD_i \phi^\alpha \cD^i \phi^{\bar{\beta}}
+2 e^K |W|^2 =0~,
\la{sfeq}
\ee
 and
\be
\label{ein3}
&&\hn_i h_j -{1 \over 2} h_i h_j -2 G_{\alpha \bar{\beta}} \cD_{(i} \phi^\alpha
\cD_{j)} \phi^{\bar{\beta}}+
2 \bigg( {1 \over 2} G_{\alpha \bar{\beta}} \cD_\ell \phi^{\alpha}
\cD^\ell \phi^{\bar{\beta}}
-{1 \over 2} e^K  G^{\alpha \bar{\beta}} D_\alpha W D_{\bar{\beta}} \bar{W}
\cr
 &&\qquad \qquad +e^K |W|^2 \bigg) \delta_{ij}=0~.
\ee
In what follows, we shall use the conditions implied by the KSE and the Einstein equation, and the compactness of ${\cal S}$ to show that ${\cal S}$
is topologically $T^2$.

\subsubsection{Gauge Field equations}

The gauge field equations are
\be
\nabla^M \bigg( {\rm Re} (H_{ab}) F^b{}_{MN} \bigg)
+ f_{abc} F^b{}_{MN} A^{c M}
+ G_{\alpha \bar{\beta}} \xi^\alpha_a \cD_N \phi^{\bar{\beta}}
+G_{\bar{\alpha} \beta} \xi^{\bar{\alpha}}_a \cD_N \phi^{\beta}=0~,
\la{gfe}
\ee
where $f_{abc} =  {\rm Re} (H_{ad}) f^d{}_{bc}$, and
we have neglected the contribution from the Chern-Simons term,
which vanishes for the magnetically charged near-horizon solutions.

As the sign of the flux $F^a$ depends on whether $\tau_{L \pm} \neq 0$, as a consequence of
({\ref{alg4}}), one finds that if $\tau_{L \pm \neq 0}$  then on substituting the remaining
conditions on the near-horizon geometry into the gauge field equations, one obtains
\be
\label{geq}
\mu_a h_i - \partial_\alpha \mu_a \cD_i \phi^\alpha - \partial_{\bar{\alpha}} \mu_a \cD_i \phi^{\bar{\alpha}}
\pm i \epsilon_i{}^j \big( \partial_\alpha \mu_a \cD_j \phi^\alpha - \partial_{\bar{\alpha}} \mu_a \cD_j \phi^{\bar{\alpha}} \big)
=0
\ee
where we note that in the holomorphic basis, $\epsilon_{1 \bar{1}}=-i$.

\subsubsection{Scalar Field Equations}

The scalar field equations are
\bea
G_{\alpha \bar{\beta}} \big( \nabla_M \cD^M \phi^{\bar{\beta}}
- (\partial_{\bar{\rho}} \xi^{\bar{\beta}}_a )A^a_M \cD^M \phi^{\bar{\rho}} \big)
+ \partial_{\bar{\rho}} G_{\alpha \bar{\beta}} \cD_M \phi^{\bar{\rho}}
\cD^M \phi^{\bar{\beta}}
\nn
-{1 \over 4} \big( \partial_\alpha {\rm Re} (H_{ab}) \big) F^a{}_{MN} F^{b MN}
-{1 \over 2} \partial_\alpha \big( {\rm Re} (H^{ab}) \mu_a \mu_b \big)
\nn
- \partial_\alpha \bigg( e^K \big( G^{\sigma \bar{\beta}} D_\sigma W D_{\bar{\beta}} {\bar{W}} -3 |W|^2 \big) =0
\eea
where we have neglected the contribution from the Chern-Simons term, which  vanishes
for the magnetically charged near-horizon solutions under consideration here.
On substituting the remaining
conditions on the near-horizon geometry into the scalar field equations, one obtains
\bea
\label{sceq}
G_{\alpha \bar{\beta}} \big(\hn_i \cD^i \phi^{\bar{\beta}} -h^i \cD_i \phi^{\bar{\beta}}
- \partial_{\bar{\sigma}} \xi^{\bar{\beta}}_a A^a_i \cD^i \phi^{\bar{\sigma}} \big)
+ \partial_{\bar{\sigma}} G_{\alpha \bar{\beta}} \cD_i \phi^{\bar{\sigma}} \cD^i \phi^{\bar{\beta}}
- \mu^a \partial_\alpha \mu_a
\nn
- \partial_\alpha \bigg( e^K \big( G^{\sigma \bar{\beta}} D_\sigma W D_{\bar{\beta}} {\bar{W}} -3 |W|^2 \big)
\bigg) =0
\eea

\newsection{Horizon topology}

To establish that the topology of ${\cal S}$ is $T^2$ several special cases have to be considered.  We shall begin
by assuming that $\tau_{L+}\not=0$ and $\tau_{L-}\not=0$ .

\subsection{ Solutions with $\tau_{L+}\not=0$ and $\tau_{L-}\not=0$}

If $\tau_{L+}\not=0$ and $\tau_{L-}\not=0$, then the gaugino KSE and the integrability conditions (\ref{incon}) imply that
\bea
F^a=\mu^a=dV=0~.
\eea
Moreover, (\ref{incon}) can be written as
\bea
 \tilde R_{{\cal S}}= 2\Delta + \tilde\nabla^k h_k ~.
\la{inconb}
\eea
Since $\Delta\geq 0$, the Euler number of ${\cal S}$ is not negative and so ${\cal S}$ is topologically either $T^2$ or $S^2$.

Now if $\Delta=0$, the Euler number of ${\cal S}$ vanishes and so ${\cal S}$ is topologically $T^2$. It remains
to investigate the case  $\Delta>0$.   Since the field strength $F^a=0$ and ${\cal S}$ is topologically $S^2$, the
gauge connection is trivial and so we set $A=0$. As we have mentioned the Euler number is positive. So  to avoid contradiction with
(\ref{sfeq}),  some of the scalars $\phi$ must have a {\it non-trivial} dependence on the coordinates of ${\cal S}$.

The dependence of $\phi$ on the coordinates of ${\cal S}$ is restricted by supersymmetry.
In particular from the results of \cite{class2},
\bea
i_Z d\phi=0~,
\la{conz}
\eea
where $Z$ is the 1-form bilinear, see appendix A.
 $\phi$ depends only on the coordinates of ${\cal S}$ but the components of $Z$ have $u$ and $r$ polynomial dependence. As a result
 (\ref{conz}) gives rise to a system of conditions on $\phi$, one for every polynomial $u,r$-component of $Z$. If two such conditions
 are linearly independent, (\ref{conz}) would imply that $\phi$ is constant leading to a contradiction.
 So to maintain
that some of the scalars have a non-trivial dependence on the coordinates of ${\cal S}$,
 all $u,r$ components of $Z$ along ${\cal S}$ must
be linearly dependent. A straightforward calculation reveals that this is the case provided that
\be
\label{recond}
a \bigg({1 \over 2} {\bar{b}} h_\bo -{i \over \sqrt{2}} e^{K \over 2} {\bar{W}} b \bigg)
\ee
is a {\it real} valued function.
Since $a \neq 0$ and $b \neq 0$,  ({\ref{recond}}) together with
({\ref{covdx}}) imply that
\be
a^{-1} \bigg( {1 \over 2} b h_1 +{i \over \sqrt{2}} e^{K \over 2} W {\bar{b}} \bigg)
\ee
is a real {\it constant}. As
\be
\eta_{L+} = \bigg( u ({1 \over 2} b h_1 +{i \over \sqrt{2}} e^{K \over 2} W {\bar{b}}) + a \bigg)1
\ee
it follows that we may, without loss of generality, set $a=0$, i.e. $\tau_{L+}=0$,
by making a co-ordinate transformation of the form $u=u'+c$ for an appropriately chosen
real constant $c$. Note that this transformation preserves the form of the near-horizon metric. Therefore
for $\Delta\not=0$, it suffices to consider that either $\tau_{L+}$ or $\tau_{L-}$ vanishes.

\subsection{Solutions for either $\tau_{L+}=0$ or  $\tau_{L-}=0$ and $\Delta\not=0$}

To investigate this case define
\be
\label{kkv1}
\kappa = \begin{cases} -{1 \over 2}|b|^2 h -{i \over \sqrt{2}} e^{K \over 2} W {\bar{b}}^2 \bbe^1
+ {i \over \sqrt{2}} e^{K \over 2} \bar{W} b^2 \bbe^\bo, & \mbox{if } \tau_{L-} \neq 0 \\
\Delta^{-1} \big({1 \over 2} |a|^2 h +{i \over \sqrt{2}} e^{K \over 2} {\bar{W}} a^2 \bbe^1
-{i \over \sqrt{2}} e^{K \over 2} W {\bar{a}}^2 \bbe^\bo \big), & \mbox{otherwise}~. \end{cases}
\ee
Observe that the components of $\kappa$ correspond to the $u^1 r^0$ component of
$Z_1, Z_\bo$ from ({\ref{zcomp}}) if $\tau_{L-} \neq 0$, and an appropriately chosen re-scaling of the
$u^0 r^1$ component of $Z_1, Z_\bo$ if $\tau_{L-}=0$, respectively. Moreover,
by construction, $\kappa$ is not identically zero.
As $Z$ is a Killing vector field on the near horizon geometry,
one can show  that $\kappa$ is an {\it isometry} of the horizon section ${\cal{S}}$.

To continue consider first the case $\tau_{L+}=0, \tau_{L-} \neq 0$. Note in particular that
\bea
i_{\partial \over \partial u} Z &=& 2 \sqrt{2} (1+{1 \over 2} \Delta u r)^2 |b|^2
-\sqrt{2} r^2 u^2 \Delta \big(|b|^2 (-{1 \over 8}h^2 -{1 \over 2} e^K |W|^2)
\nn
&+&{i \over 2 \sqrt{2}} b^2 e^{K \over 2} {\bar{W}} h_1 -{i \over 2 \sqrt{2}} {\bar{b}}^2
e^{K \over 2} W h_\bo \big)
\eea
As $Z$ is a smooth 1-form and ${\partial \over \partial u}$ is a smooth Killing vector field
in the near-horizon spacetime, it follows that this scalar is also smooth. On evaluating it
on the horizon section, $r=u=0$, it follows that $|b|^2$ is a smooth function on ${\cal{S}}$.
In addition,  as $\kappa$ is obtained from the pull-back
of $\cL_{\partial \over \partial u} Z$ on ${\cal S}$, it follows that $\kappa$ is
a smooth 1-form on ${\cal{S}}$.
Moreover, ({\ref{covdx}}) implies that
\be
\label{alt1}
 d(\Delta |b|^2)=- \Delta \kappa~.
\ee
As $\kappa$ is Killing, we further find
\be
\label{alt2}
\hn_i \hn_j (\Delta |b|^2) = \Delta^{-1} \hn_{(i} \Delta \hn_{j)} (\Delta |b|^2)~.
\ee
Assuming that ${\cal{S}}$ is not topologically $T^2$, and so ${\cal{S}}$ has a non-vanishing Euler number, it follows
that $\kappa$ must vanish at some point $P \in {\cal{S}}$. Using (\ref{alg1}), this implies that $b=0$ at $P$.
Furthermore ({\ref{alt1}}) and ({\ref{alt2}}) then imply that all covariant derivatives of
$\Delta |b|^2$ must also vanish at $P$.  Assuming that $\Delta |b|^2$ is analytic\footnote{This result may hold for  $\Delta |b|^2$
 smooth but we have not been able to extend the proof.} on ${\cal{S}}$,  it follows that
$b=0$ everywhere. However, this leads immediately to a contradiction, as it implies that the Killing
spinor must vanish everywhere.

Next, consider the case $\tau_{L+} \neq 0$, $\tau_{L-}=0$. In this case, we have
\bea
i_{\partial \over \partial r} Z &=& - 2 \sqrt{2}  |a|^2~.
\eea
As $Z$ is a smooth 1-form and $\partial \over \partial r$ is a smooth
vector field
in the near-horizon spacetime, it follows that $|a|^2$ is a smooth function on ${\cal{S}}$.
In addition, as $\kappa$ is obtained from a linear combination of the
pull-back of $\cL_{\partial \over \partial r} Z$ to $u=r=0$ and $|a|^2 h$, it follows that
$\kappa$ is a smooth 1-form on ${\cal{S}}$.
Moreover, ({\ref{covdx}}) implies that
\be
\label{alt3}
d(|a|^2)= \Delta \kappa~.
\ee
As $\kappa$ is Killing, we further find
\be
\label{alt4}
\hn_i \hn_j (|a|^2) = \Delta^{-1} \hn_{(i} \Delta \hn_{j)} (|a|^2)~.
\ee
Repeating a similar argument to the one we have used for the previous case, one finds that, unless ${\cal{S}}$ is topologically $T^2$,
$\tau_{L+} = 0$. Thus if $\Delta\not=0$, then one concludes that ${\cal{S}}$ is topologically $T^2$.

\subsection{Solutions for either $\tau_{L+}=0$ or  $\tau_{L-}=0$ and $\Delta=0$}

There are two separate cases to consider depending on whether the components of $Z$ along ${\cal S}$ vanish or not.
To continue denote  the components of $Z$ along ${\cal S}$ with $\tilde Z$.

\subsubsection{$\tilde Z\not=0$}

In this case observe that  the supersymmetry conditions (\ref{susycona})
can be rewritten as
\bea
i_ZF^a=i_{\tilde Z} F^a=0~,~~~i_Z \cD \phi^\alpha=i_{\tilde Z} \cD \phi^\alpha=0~.
\eea
Since $F^a$ has one non-vanishing spacetime component, if $\tilde Z\not=0$, then one concludes that
\bea
F^a=0~.
\eea
Similarly, if $\tilde Z\not=0$, one concludes that the scalars depend on only one coordinate and so
\bea
dV=0~.
\eea
Using the integrability conditions (\ref{incon}) for  either $\tau_{L+}$ or  $\tau_{L-}$, one finds that
\bea
\tilde R_{\cal S}=\tilde \nabla^k h_k~.
\eea
Thus the Euler number of ${\cal S}$ vanishes and so ${\cal S}$ is topologically $T^2$.

\subsubsection{Solutions with $\tau_{L+}=0$ and  $\tilde Z =0$}

Using $\tilde Z =0$, one finds that the vector field associated to the 1-form bilinear ({\ref{zcomp}}) is
\be
Z = 2 \sqrt{2} |b|^2 {\partial \over \partial r}~.
\ee
Note that as $g(Z,{\partial \over \partial u})= 2 \sqrt{2} |b|^2$, it follows that
$|b|^2$ is a smooth function on ${\cal{S}}$. Furthermore,
the requirement that  $Z$ is Killing gives
\be
\label{yu2}
|b|^2 h + d |b|^2 =0~.
\ee
This condition implies that if $b=0$ at any point in ${\cal{S}}$, then $b=0$ everywhere.
Hence, for the solution to be supersymmetric, we take $b \neq 0$ everywhere.
Next, note that ({\ref{alge2}}) can be rewritten as
\be
\hn_i \hn^i (|b|^{-1}) = |b|^{-1} e^K G^{\alpha \bar{\beta}} D_\alpha W
D_{\bar{\beta}} {\bar{W}} \ .
\ee
On integrating this expression over ${\cal{S}}$, the contribution from the LHS vanishes,
and one obtains the conditions
\be
h=0, \qquad D_\alpha W=0, \qquad W=0 \ ,
\la{rect}
\ee
and $|b|$ is a nonzero constant.

Using (\ref{rect}), (\ref{alg6}) implies that
\bea
\label{nx1}
\cD_1\phi^{\bar \alpha}=0~,
\eea
ie the scalar fields up to a gauge transformation are holomorphic.
In turn, the gauge equation ({\ref{geq}}) implies that
\be
\label{nx2}
\partial_\alpha \mu_a \cD_i \phi^\alpha =0 \ ,
\ee
and the scalar equation ({\ref{sceq}}) implies
\be
\label{nx3}
\mu^a \partial_\alpha \mu_a=0 \ .
\ee
Furthermore (\ref{sfeq}) implies
\be
\tilde R_{{\cal S}} - 2 \mu_a \mu^a -2 G_{\alpha \bar{\beta}} \cD_1 \phi^\alpha \cD_{\bar 1} \phi^{\bar{\beta}}=0~.
\la{sfeq1}
\ee
Thus provided that inner product  ${\rm Re} H$,   (\ref{gfe}), of the gauge group and the K\"ahler metric  are positive
definite,  the Euler number
of ${\cal S}$ is non-negative. This is a mild assumption on the couplings.
Therefore ${\cal S}$ is topologically either $T^2$ or $S^2$.
In particular ${\cal S}$ is topologically $T^2$ if, and only if, $\mu^a=0$
and $\cD \phi^\alpha=0$, ie if and only if the gauge field vanishes and the scalars are constant. In turn this
implies that ${\cal S}$ is isometric to $T^2$.
For all these horizons $\Delta=h=0$, hence the near horizon geometry is
$\bR^{1,1}\times T^2$ or $\bR^{1,1}\times S^2$ though the metric on $S^2$ may not be the round
one.

\subsubsection{Solutions with $\tau_{L-}=0$ and  $\tilde Z =0$}

Using $\tilde Z =0$, one finds that the vector field associated to the 1-form bilinear ({\ref{zcomp}}) is
\be
Z = -2 \sqrt{2} |a|^2 {\partial \over \partial u}~.
\ee
Moreover $|a|^2$ is {\it constant} because  $Z$ is  Killing. Thus $a\not=0$ everywhere on the spacetime.
Note that the condition ${\tilde Z}=0$ implies that
\be
{1 \over 2} |a|^2 h_1 +{i \over \sqrt{2}} e^{K \over 2} {\bar{W}} a^2 =0~,
\la{xx2}
\ee
and the chiral KSE conditions ({\ref{alg6}}) and ({\ref{alg7}}) are equivalent to
\be
\sqrt{2} i {\bar{a}} \cD_1 \phi^\alpha -a e^{K \over 2} G^{\alpha \bar{\beta}} D_{\bar{\beta}} {\bar{W}}=0~.
\la{xx3}
\ee
In addition, the gauge field equation ({\ref{geq}}) gives
\be
\label{eo1}
\mu_a h_1 - 2 \partial_{\bar{\alpha}} \mu_a \cD_1 \phi^{\bar{\alpha}}=0~.
\ee
On comparing ({\ref{alg3}}) with ({\ref{ein3}}) and after some computation,  one obtains the condition
\be
W A^a_i \big( \xi^\alpha_a \partial_\alpha K + \xi^{\bar{\alpha}}_a \partial_{\bar{\alpha}} K \big) =0~.
\ee
Observe that the above equation is satisfied provided that the K\"ahler potential is invariant under the
isometries generated by $\xi$.

Next observe that (\ref{xx2}) and (\ref{xx3}) imply that
\bea
{\cal D}_{\bar1} \phi^{\bar\a}-{1\over2} h_{\bar 1} W^{-1} G^{\bar\a\b} D_\b W=0~.
\eea
Using this and other conditions derived from the KSEs, one can show after some computation that
  the scalar equation ({\ref{sceq}}) can be rewritten as
\be
\label{eo2}
\mu^a \partial_\alpha \mu_a =0~.
\ee
On combining ({\ref{eo1}}) with ({\ref{eo2}}), one obtains
\be
\mu^a \mu_a h =0~.
\ee
Since the kinetic term of the gauge fields is canonical, ie the inner product ${\rm Re} H$ in   (\ref{gfe}) is positive
definite, either $\mu^a=0$ for all $a$,
or $h=0$.

Suppose first that there exists some $\mu^a \neq 0$. Then $h=0$ implies
the conditions
\be
W=0, \qquad D_\alpha W =0~,
\ee
and the conditions from the chiral, gauge and scalar field and KSEs simplify to
\be
\cD_1 \phi^\alpha=0, \qquad \partial_{\alpha} \mu_a \cD_i \phi^{\alpha}=0,
\qquad \mu^a \partial_\alpha \mu_a =0 \ .
\ee
These conditions are identical to those derived in the previous section with the only difference that
here the scalar fields, up to a gauge transformation, are antiholomorphic instead of holomorphic.
As a result (\ref{sfeq}) implies
\be
\tilde R_{{\cal S}} - 2 \mu_a \mu^a -2 G_{\alpha \bar{\beta}} \cD_{\bar 1} \phi^\alpha \cD_{ 1} \phi^{\bar{\beta}}=0~.
\la{sfeq1x}
\ee
As in the previous case provided that inner product  ${\rm Re} H$,   (\ref{gfe}), of the gauge group and the K\"ahler metric  are positive
definite,  the Euler number
of ${\cal S}$ is non-negative.
Therefore ${\cal S}$ is topologically either $T^2$ or $S^2$.
In particular ${\cal S}$ is topologically $T^2$ if, and only if, $F=\mu=0$ and the scalars are constant.
In turn this implies that ${\cal S}$ is isometric to $T^2$.
Again for all these horizons $\Delta=h=0$, and so the near horizon geometry is
$\bR^{1,1}\times T^2$ or $\bR^{1,1}\times S^2$ though the metric on $S^2$ may not be the round
one.

Alternatively, suppose that $\mu^a=0$ for all $a$. Then $F^a=0$ for all $a$, and under the assumption we have made
one can without loss of generality work locally in a gauge for which $A^a=0$.
Note that in such a case, the metric and scalars satisfy the field equations obtained from
coupling gravity to scalar matter with a scalar potential. In particular, this type
of Lagrangian was considered in the analysis of \cite{hollands}, in which it was shown that
if one assumes that the spacetime metric and the scalars are analytic, then the
horizon section ${\cal{S}}$ admits a rotational isometry $\kappa$.
If we assume sufficient conditions on the metric and scalars such that the rigidity theorem
of \cite{hollands} holds here \footnote{It is not a priori clear that
the scalars are globally well-defined and analytic.
However, in the analysis of \cite{hollands}, it appears that only analyticity of
the metric is required in order to construct a preferred set of Gaussian Null co-ordinates.},
then one finds that the Lie derivative of the scalars with respect to
$\kappa$ vanishes; i.e. the scalars depend only on one co-ordinate.
This in turn implies that $dV=0$, and hence from ({\ref{incon}}) it follows that
the Euler number of ${\cal{S}}$ vanishes, so ${\cal{S}}$ is topologically $T^2$.

\vskip 0.5cm
\noindent{\bf Acknowledgements} \vskip 0.1cm
\noindent
We thank  P~Meessen and T~ Ortin who alerted  us to the existence of spherical horizons
excluded in an earlier version of this work  because of an additional assumption we had made.
JG is supported by the EPSRC grant, EP/F069774/1. GP is
partially supported
by the EPSRC grant EP/F069774/1 and the STFC rolling grant ST/G000/395/1.
\vskip 0.5cm

 \setcounter{section}{0}

\appendix{Conventions and bilinear}

The non-vanishing components of the spin connection associated with the basis ({\ref{basis1}}) are
\bea
&&\Omega_{-,+i} = -{1 \over 2} h_i~,~~
\Omega_{+,+-} = -r \Delta, \quad \Omega_{+,+i} =r^2( {1 \over 2}  \Delta h_i -{1 \over 2} \partial_i \Delta),
\cr
&&\Omega_{+,-i} = -{1 \over 2} h_i, \quad \Omega_{+,ij} = -{1 \over 2} r dh_{ij}~,~~~\Omega_{i,+-} = {1 \over 2} h_i~,
\cr
&& \quad \Omega_{i,+j} = -{1 \over 2} r dh_{ij}~,~~~\Omega_{i,jk}= \tilde\Omega_{i,jk}~,
\eea
where $\tilde\Omega$ denotes the spin-connection of the horizon section ${\cal S}$ in the $\bbe^i$ basis.

To find the Killing spinor bilinear form $Z$, we use $
\Gamma_{12} 1 = e_{12}$, $\Gamma_{12} e_1 = - e_2$ and $\Gamma_{12} e_2 = e_1$, and (\ref{aba}). Then,
one  obtains
\bea
\label{zcomp}
Z_+ &=& {1 \over \sqrt{2}} r^2 \big(|a|^2 ({1 \over 2} h^2+2 e^K |W|^2)
+ \sqrt{2}i a^2 h_\bo e^{K \over 2} {\bar{W}} - \sqrt{2}i {\bar{a}}^2 h_1 e^{K \over 2} W \big)
\nn
&+&\sqrt{2} r \big(b(-{\bar{a}} h_1 - \sqrt{2} i e^{K \over 2} {\bar{W}} a)
+ {\bar{b}} (-a h_\bo + \sqrt{2} i e^{K \over 2} W {\bar{a}}) \big) + 2 \sqrt{2} |b|^2
\nn
&+& 2 \sqrt{2} u \bigg({1 \over 4}  r^2 \Delta \big(b(-{\bar{a}} h_1 - \sqrt{2} i e^{K \over 2} {\bar{W}} a)
+ {\bar{b}} (-a h_\bo + \sqrt{2} i e^{K \over 2} W {\bar{a}})\big) + r \Delta |b|^2 \bigg)
\nn
&+& {1 \over \sqrt{2}} r^2 \Delta^2 |b|^2 u^2
\nn
Z_- &=&-2 \sqrt{2} |a|^2  + 2 \sqrt{2} u \bigg(-{1 \over 2} a \bar{b} h_\bo -{1 \over 2} \bar{a} b h_1
+{i \over \sqrt{2}} a e^{K \over 2} {\bar{W}} b -{i \over \sqrt{2}} \bar{a} e^{K \over 2} W {\bar{b}} \bigg)
\nn
&+& 2 \sqrt{2} u^2 \bigg( |b|^2 (-{1 \over 8} h^2 - {1 \over 2} e^K |W|^2)
+{i \over 2 \sqrt{2}} b^2 e^{K \over 2} {\bar{W}} h_1
-{i \over 2 \sqrt{2}} {\bar{b}}^2 e^{K \over 2} W h_\bo \bigg)
\nn
Z_1 &=& -2 \sqrt{2} a \bar{b} + 2 \sqrt{2} r\big({1 \over 2} |a|^2 h_1 +{i \over \sqrt{2}} e^{K \over 2} {\bar{W}} a^2
\big)
\nn
&+& 2 \sqrt{2} u \bigg(-{1 \over 2} |b|^2 h_1 -{i \over \sqrt{2}} e^{K \over 2} W {\bar{b}}^2
\nn
&+& r \big(-{1 \over 2} \Delta a \bar{b}
+({1 \over 2} b h_1 +{i \over \sqrt{2}}e^{K \over 2} W {\bar{b}})
({1 \over 2} \bar{a} h_1 +{i \over \sqrt{2}} e^{K \over 2} \bar{W} a) \big) \bigg)
\nn
&+& \sqrt{2} r \Delta u^2 \bigg(-{1 \over 2} |b|^2 h_1 -{i \over \sqrt{2}} e^{K \over 2} W {\bar{b}}^2 \bigg)~,
\eea
and $Z_{\bar 1}=\bar Z_1$.



\begin{thebibliography}{99}

\bibitem{wess}
J. Wess and J. Bagger, \textit{Supersymmetry and supergravity,} Princeton, USA: Univ. Pr.
(1992) p259.


\bibitem{hawking}
 S.~W.~Hawking,
  ``Black holes in general relativity,''
  Commun.\ Math.\ Phys.\  {\bf 25} (1972) 152.

\bibitem{chrusc}
P. T. Chrusciel and R. M. Wald,
``On the topology of stationary black holes,"
Class. Quant. Grav. {\bf{11}} (1994) 147;  gr-qc/9410004.

\bibitem{class2}
U. Gran, J. Gutowski and G. Papadopoulos, \textit{Geometry of all supersymmetric four-dimensional N = 1 supergravity backgrounds,} JHEP 0806
(2008) 102; arXiv:0802.1779 [hep-th].

\bibitem{class3}
T. Ortin, \textit{The Supersymmetric solutions and extensions of ungauged matter-coupled N=1, d=4 supergravity,} JHEP 0805 (2008) 034;
arXiv:0802.1799 [hep-th]

\bibitem{mo}
  P.~Meessen and T.~Ortin,
  ``Ultracold horizons in gauged N=1 d=4 supergravity,''
  arXiv:1007.3917 [hep-th].

\bibitem{israel}
  W.~Israel,
  ``Event Horizons In Static Vacuum Space-Times,''
  Phys.\ Rev.\  {\bf 164} (1967) 1776.

\bibitem{carter}
  B.~Carter,
  ``Axisymmetric Black Hole Has Only Two Degrees of Freedom,''
  Phys.\ Rev.\ Lett.\  {\bf 26} (1971) 331.

\bibitem{robinson1}
  D.~C.~Robinson,
  ``Uniqueness of the Kerr black hole,''
  Phys.\ Rev.\ Lett.\  {\bf 34} (1975) 905.

\bibitem{israel2}
W. Israel, ``Event Horizons in Static, Electrovac Space-Times,"
Commun. Math. Phys. {\bf{8}} (1968) 245.

\bibitem{mazur}
P. O. Mazur, ``Proof of Uniqueness of the Kerr-Newman Black Hole Solution,"
J. Phys. A {\bf{15}} (1982) 3173.

\bibitem{robinson} D.~ Robinson, ``Four decades of black hole uniqueness theorems,'' appeared
in {\it The Kerr spacetime: Rotating black holes in General Relativity}, eds D.~L.~ Wiltshire, M.~ Visser and
S.~ M.~ Scott, pp 115-143,  CUP 2009.

\bibitem{kunduri}
  H.~K.~Kunduri, J.~Lucietti and H.~S.~Reall,
  ``Near-horizon symmetries of extremal black holes,''
  Class.\ Quant.\ Grav.\  {\bf 24} (2007) 4169
  [arXiv:0705.4214 [hep-th]].

\bibitem{hollands}
  S.~Hollands and A.~Ishibashi,
  ``On the `Stationary Implies Axisymmetric' Theorem for Extremal Black Holes
  in Higher Dimensions,''
  Commun.\ Math.\ Phys.\  {\bf 291} (2009) 403


\bibitem{schoen}
  R.~Schoen and S.~T.~Yau,
  ``Positivity Of The Total Mass Of A General Space-Time,''
  Phys.\ Rev.\ Lett.\  {\bf 43} (1979) 1457.

\bibitem{witten}
  E.~Witten,
  ``A Simple Proof Of The Positive Energy Theorem,''
  Commun.\ Math.\ Phys.\  {\bf 80} (1981) 381.


\bibitem{gibbons}
  G.~W.~Gibbons and C.~M.~Hull,
  ``A Bogomolny Bound For General Relativity And Solitons In N=2
  Supergravity,''
  Phys.\ Lett.\  B {\bf 109} (1982) 190.

  \bibitem{nonbps1}
I. Bena, G. Dall'Agata, S. Giusto, C. Ruef and  N. P. Warner,
\textit{Non-BPS Black Rings and Black Holes in Taub-NUT},
JHEP 0906 (2009) 015; arXiv:0902.4526 [hep-th].

\bibitem{nonbps2}
K. Goldstein and S. Katmadas,
\textit{Almost BPS black holes},
JHEP 0905 (2009) 058;  arXiv:0812.4183 [hep-th].

\bibitem{hethor}
J. B. Gutowski and G. Papadopoulos,
\textit{Heterotic Black Horizons}; arXiv:0912.3472 [hep-th].

\bibitem{desit}
J Grover and J. Gutowski,
\textit{Horizons in de-Sitter Supergravity;}
arXiv:1001.2460 [hep-th].

\bibitem{reallbh}
H. S. Reall,
\textit{Higher dimensional black holes and supersymmetry},
Phys. Rev. {\bf{D68}} (2003) 024024; hep-th/0211290.


























\bibitem{hawkingellis}
S. W. Hawking and G. F. R. Ellis, ``The large scale structure of space-time"
Cambridge University Press, 1973.

\bibitem{chrusc2}
P. T. Chrusciel, ``On rigidity of analytic black holes" Commun. Math. Phys. {\bf{189}} (1997) 1;
gr-qc/9610011.

\bibitem{hollands1}
S. Hollands, A. Ishibashi and R. M. Wald, ``A higher dimensional stationary rotating
black hole must be axisymmetric," Commun. Math. Phys. {\bf{271}} (2007) 699;

\bibitem{monc}
V. Moncrief and J. Isenberg, ``Symmetries of Higher Dimensional Black Holes",
Class. Quant. Grav. {\bf{25}} (2008) 195015;
[arXiv:0805.1451 [gr-qc]].













\bibitem{gnull}
H. Friedrich, I. Racz and R. M. Wald,
\textit{On the rigidity theorem for space-times with a stationary event horizon or a compact Cauchy horizon,}
Commun. Math. Phys. {\bf{204}} (1999) 691; gr-qc/9811021.





\bibitem{spingeom}
  J.~Gillard, U.~Gran and G.~Papadopoulos,
  ``The spinorial geometry of supersymmetric backgrounds,''
  Class.\ Quant.\ Grav.\  {\bf 22} (2005) 1033
  [arXiv:hep-th/0410155].




\end{thebibliography}
\end{document}